\documentclass[authoryear,10pt,letterpaper]{elsarticle}
\usepackage[left=3cm,right=3cm,top=3cm,bottom=3cm]{geometry}
\usepackage{amssymb,amsmath,verbatim,mathtools,needspace,enumitem,etoolbox,graphicx,physics,microtype,afterpage}
\usepackage{gensymb, soul}

\usepackage[dvipsnames, usenames]{xcolor}

\definecolor{linkcolor}{rgb}{0.0,0.3,0.5}
\usepackage[unicode,colorlinks=true, linkcolor=linkcolor, citecolor=linkcolor, filecolor=linkcolor,urlcolor=linkcolor]{hyperref}

\usepackage{titlesec}
\setcitestyle{aysep={}}

\usepackage{setspace}
\usepackage{tocloft}

\journal{iScience}

\begin{document}

{\LARGE \bfseries Article}\bigskip

{\LARGE Testing general relativity with gravitational-wave catalogs:}

\smallskip
{\LARGE $\,$the insidious nature of waveform systematics}
\bigskip\bigskip

{Christopher J. Moore, Eliot Finch, Riccardo Buscicchio, Davide Gerosa}

\medskip

{\small   \it
{School of Physics and Astronomy \& Institute for Gravitational Wave Astronomy, \\\phantom{$\quad\;\,$}University of Birmingham, Birmingham, B15 2TT, UK.
}}

\section*{Abstract}
\noindent
Gravitational-wave observations of binary black holes allow new tests of general relativity to be performed on strong, dynamical gravitational fields.
These tests require accurate waveform models of the gravitational-wave signal, otherwise waveform errors can erroneously suggest evidence for new physics. 
Existing waveforms are generally thought to be accurate enough for current observations, and each of the events observed to date appears to be individually consistent with general relativity.
In the near future, with larger gravitational-wave catalogs, it will be possible to perform more stringent tests of gravity by analyzing large numbers of events together.
However, there is a danger that waveform errors can accumulate among events: even if the waveform model is accurate enough for each individual event, it can still yield erroneous evidence for new physics when applied to a large catalog.
This paper presents a simple linearised analysis, in the style of a Fisher matrix calculation, that reveals the conditions under which the apparent evidence for new physics due to waveform errors grows as the catalog size increases.
We estimate that, in the worst-case scenario, evidence for a deviation from general relativity might appear in some tests using a catalog containing as few as $10\text{--}30$ events above a signal-to-noise ratio of 20. 
This is close to the size of current catalogs and highlights the need for caution when performing these sorts of experiments.

%\tableofcontents
%%%
\section{Introduction} \label{sec:intro}

The detection of gravitational waves (GWs)  by LIGO~\citep{2015CQGra..32g4001L} and Virgo \citep{2015CQGra..32b4001A} has made possible new tests of general relativity (GR) in the strong-field regime \citep{2013LRR....16....9Y,2015CQGra..32x3001B,2016PhRvD..94h4002Y,2019PhRvD.100j4036A, 2020arXiv201014529T}. 
Numerous detections of binary coalescences have been made so far (mostly binary black holes) and in the coming years the size of this catalog of detections will continue to grow \citep{2018LRR....21....3A}.
Furthermore, observations of GWs in different frequency bands will reveal new types of sources and enable other, complementary tests of GR. 
As the sensitivity of the instruments improve, the signal-to-noise ratios (SNR) of the loudest individual events will increase and such tests will become increasingly stringent.

However, care must be taken when interpreting the results of these test or we risk incorrectly claiming evidence for new physics.

Among the wide range of possible tests of GR are parametric tests. These involve the introduction of additional degrees of freedom to the theory, which are described by one or more new parameters. These additional quantities are then measured along with the astrophysical source parameters. Frameworks for performing parametric tests include both the introduction of artificial coefficients to various terms in the waveform (for a review, see \citealt{2014LRR....17....4W}) and extensions involving specific beyond-GR theories, such as those motivated by quantum gravity \citep{2016PhRvD..94h4002Y}. (See \citealt{2020arXiv200608918C} for a discussion of possible drawbacks of parametric tests.) 

As most GW signals will have a SNR close to detection threshold \citep{2011CQGra..28l5023S,2014arXiv1409.0522C}, combining information from multiple signals is an attractive %\cm{I didn't like ``precious''} 
avenue to perform stronger tests.
The way different events are put together crucially depends on the test one wishes to perform \citep{2019PhRvD..99l4044Z}. 
For GR modifications with parameters that are thought to be common among all the events in the catalog (for instance, the mass of the graviton), one should multiply the individual likelihoods on the deviation parameters to find the combined, catalog-level likelihood on the deviation. 
If instead each event can have a different, independent deviation parameter value (for instance, addition black hole degrees of freedom, aka \emph{hairs})
one should instead find the combined, catalog evidence by multiplying the individual event Bayes' factors. 
These possibilities represent the two extrema of a broader class of catalog tests where the non-GR parameters follow some distribution that can depend on the GR quantities (such as masses and spins), and can be tackled using a hierarchical Bayesian approach \citep{2019PhRvL.123l1101I}. Some of the analyses presented by \cite{2020arXiv201014529T} were carried out within this framework.

Tests of GR can be affected by inaccuracies in the GW signal models used to analyse the data~ \citep{2008PhRvD..78l4020L,2020PhRvR...2b3151P}.
Waveform systematics can erroneously lead to evidence for a deviation from GR. In other words, even if GR is the correct description of nature, unmodelled waveform systematics can lead us to believe the opposite.
Going beyond single events, in this paper we explore the role of waveform systematics when using multiple signals in a catalog to test GR.
For each event, we employ a simplified linear analysis, similar to that by \cite{2007PhRvD..76j4018C,2013PhRvD..87j2002V,2015PhRvD..91l4062G},
and study how systematic waveform errors introduce biases in the beyond GR parameters.
By studying the two extreme cases highlighted above ---GR deviations which are common and different among events--- we extend the linear analysis to show how the effects of waveform systematics can accumulate as the size of the catalog grows. 
Even if the imperfect waveform model is good enough to safely analyse each of the events individually, it may induce evidence for beyond GR physics with arbitrarily high confidence when applied to a large catalog.
This serves to highlight the dangerous and insidious nature waveform systematics in GW catalogs.

This paper is organised  as follows. In Sec.~\ref{sec:linear_sig_analysis}, we derive the building blocks of our analysis by investigating the impact of systematics on single events. In Sec.~\ref{sec:catalog}, we illustrate how those ingredients enter a  catalog analysis. In Secs.~\ref{sec:toy} and \ref{sec:real}, we present the results of our findings applied to two sets of simulated event catalogs of increasing complexity.
Finally, in Sec.~\ref{sec:discussion}, we draw our conclusions.

%%%
\section{Linear signal analysis: single events} \label{sec:linear_sig_analysis}

In order to establish how the evidence for new physics scales with the individual event SNR and the number of events in the catalog, we use a simplified, linearised signal analysis, in the spirit of a Fisher matrix calculation.

In the following analysis, it is assumed that GR is the correct description of nature.
For each individual GW event the observed data contains a sum of instrumental noise, $n$, and a GW signal.
Here, we restrict our analysis to the case when a single interferometer is involved in the observation; the extension to multiple interferometers is straightforward and does not significantly change the following arguments.
The observed data, $s$, can be written
\begin{align} \label{eq:obs_data}
    s = n + \underbrace{h(\alpha_\mathrm{Tr}=0;\theta_{\rm Tr}) + \Delta h(\theta_{\rm Tr})}_{\textrm{GW Signal}}\,.
\end{align}
The true signal is (hopefully) close to our waveform model, $h(\alpha=0;\theta)$, but will inevitably include some modelling error, denoted here by $\Delta h(\theta)$.
Our model, $h(\alpha;\theta)$, is a function of the parameters that describe the source in GR, $\theta^i$. This includes both intrinsic (masses, spins, etc.) and extrinsic (sky position, distance, etc.) quantities.
The true source parameters (which are unknown \emph{a priori}) are denoted by $\theta_{\rm Tr}$.
The model error is also a function of $\theta$ 
(for example, regions of parameter space with asymmetric mass ratios and strong spin precession will typically have larger model errors; 
\citealt{2020PhRvR...2b3151P}).
As we are searching for parameterised deviations from GR, our model, $h(\alpha;\theta)$, is also a function of at least one modified gravity parameter, $\alpha$.
This parameter quantifies the deviation from GR and we assume it is defined such that GR is smoothly recovered in the limit $\alpha\rightarrow 0$.

The analysis of the single GW event in Eq.~(\ref{eq:obs_data}) 
involves performing Bayesian parameter inference over the combined parameter space ${\lambda}\equiv(\alpha;\theta)$.
Assuming the instrumental noise is Gaussian, the likelihood $\mathcal{L}(\alpha;\theta)\equiv P(s|\alpha;\theta)$ is given by 
\begin{align} \label{eq:like}
    \log \mathcal{L}(\alpha;\theta) &= -\frac{1}{2}\left|s-h(\alpha;\theta)\right|^2 + c\,, \\
    &=-\frac{1}{2}\left|n-\delta h(\alpha;\theta)+\Delta h(\theta_{\rm Tr})\right|^{2} + c \,. \nonumber
\end{align}
The norm is defined as $|\cdot|^{2}=\left<\cdot|\cdot\right>$, where angle brackets denote the usual signal inner product \citep{1987thyg.book..330T,2015CQGra..32a5014M}.
The constant $c$ is an unimportant normalization, and we have defined $\delta h(\alpha;\theta)\equiv h(\alpha;\theta)-h(\alpha_{\rm Tr};\theta_{\rm Tr})$, where $\theta_{\rm Tr}$ and $\alpha_{\rm Tr}=0$ denote the true parameters.
For simplicity, we assume that the prior on $\lambda$ is approximately flat within the range that $\mathcal{L}$ has significant support; therefore, the likelihood is proportional to the Bayesian posterior distribution.

If the SNR is large, then the posterior is expected to be strongly peaked in a relatively narrow region around $\theta_{\rm Tr}$ and $\alpha_{\rm Tr}$.
We assume that this region is small enough that the model can be approximated as being linear in all parameters. 
We Taylor-expand our model about the true parameters as follows;
\begin{align} \label{eq:linear_model}
    \delta h(\alpha;\theta) &\approx \frac{\partial h}{\partial \alpha}\bigg|_{(0,\theta_{\rm Tr})}\alpha + \frac{\partial h}{\partial \theta^{\,i}}\bigg|_{(0, \theta_{\rm Tr})}\delta\theta^{\,i} +\ldots \\
    &= \frac{\partial h}{\partial \lambda^{\mu}}\bigg|_{\lambda_{\rm Tr}}\delta\lambda^{\mu}+\mathcal{O}(\delta\lambda^2)\,. \nonumber
\end{align}
We have defined $\delta \theta = \theta-\theta_{\rm Tr}$, and $\delta \lambda = \lambda-\lambda_{\rm Tr}$ and all derivatives are evaluated at the true parameters (hereafter, this will be omitted from our notation).
The index $\mu$ labels the components of the combined parameter vector, $\lambda^\mu$.
Hereafter, we retain only leading order terms in $\delta{\lambda}$.

The likelihood in Eq.~(\ref{eq:like}) is peaked at the maximum likelihood (ML) parameters, $\lambda_{\rm ML}$, which are defined implicitly by
\begin{align} \label{eq:MLdef}
    \frac{\partial \log\mathcal{L}}{\partial \lambda}\bigg|_{\lambda=\lambda_{\rm ML}} = 0\,.
\end{align}
Using Eqs.~(\ref{eq:like}-\ref{eq:linear_model}), this can be solved to find 
\begin{align}\label{eq:MLparams}
    {\lambda}_{\rm ML} = {\lambda}_{\rm Tr} + \Delta{\lambda}_{\rm stat}+\Delta{\lambda}_{\rm sys} \,, 
\end{align}
where (using Einstein summation convention)
\begin{align}
    \Delta{\lambda}_{\rm stat}^{\mu} &=  \left(\Gamma^{-1}\right)^{\mu\nu}
    \left<n\Big|\frac{\partial h}{\partial{\lambda}^{\nu}}\right> \,, \label{eq:dn}\\
    \Delta{\lambda}_{\rm sys}^{\mu} &= \left(\Gamma^{-1}\right)^{\mu\nu}\left<\Delta h({\theta}_{\rm Tr})\Big|\frac{\partial h}{\partial{\lambda}^{\nu}}\right> \,, \label{eq:dsys}
\end{align}
and $\Gamma_{\mu\nu}$ is the Fisher matrix, 
\begin{align}\label{eq:Fisher}
     \Gamma_{\mu\nu} = \left< \frac{\partial h}{\partial{\lambda}^{\mu}} \Big| \frac{\partial h}{\partial{\lambda}^{\nu}} \right> \,.
\end{align}
From Eq.~(\ref{eq:MLparams}), it can be seen that the ML parameters are close to the true source parameters, but shifted by both statistical and systematic errors.
The statistical error $\Delta \lambda_{\rm stat}$ depends on the random noise realisation, $n$, in the observed data.
The systematic error $\Delta \lambda_{\rm sys}$ depends on the model error $\Delta h$.

Now that we have found the location of the maximum likelihood, we may evaluate the second derivatives of Eq.~(\ref{eq:like}), $\partial_{\mu}\partial_{\nu}\log\mathcal{L}$, at the ML parameters and expand the log-likelihood to second order about this point.
Doing this, we find
\begin{align} \label{eq:full_posterior}
    \log\mathcal{L}({\lambda}) \approx c' - \frac{1}{2}\Gamma_{\mu\nu}\left({\lambda}-{\lambda}_{\rm ML}\right)^{\mu}\left({\lambda}-{\lambda}_{\rm ML}\right)^{\nu} \,,
\end{align}
where $c'$ is another unimportant normalization constant. 
Within the approximations that have been made, the likelihood (and the posterior) is approximately a multivariate Gaussian on the parameters $\lambda$ with mean vector ${\lambda}_{\rm ML} \equiv(\alpha_{\rm ML}, {\theta}_{\rm ML})$ and covariance matrix $\Gamma^{-1}$.

We wish to use the observed data to test GR. 
Therefore, we investigate the 1D marginalised posterior on the $\alpha$ parameter to see if it is peaked away from the GR value, $\alpha=0$. 
Because the full posterior in Eq.~(\ref{eq:full_posterior}) is a multivariate Gaussian, the 1D marginalization integral can be carried out analytically.
The 1D marginalised posterior on the $\alpha$ parameter reads
\begin{align} \label{eq:1Ddist}
    P(\alpha)= \int\mathrm{d}\theta\;\mathcal{L}({\theta},\alpha) = \frac{\exp\left[-\frac{
     (\alpha-\alpha_{\rm ML})^{2}}{2\sigma_\alpha^{2}}
     \right]}{\sqrt{2\pi \sigma_\alpha^{2}}} \,, 
\end{align}
where $\alpha_{\rm ML}=\alpha_{\rm stat}+\alpha_{\rm sys}$, and
\begin{align}
    \alpha_{\rm stat} &= \left(\Gamma^{-1}\right)^{0 \nu} 
    \left<n\Big|\frac{\partial h}{\partial{\lambda}^{\nu}}\right> \,, \label{eq:AlphaStat0}\\
    \alpha_{\rm sys} &= \left(\Gamma^{-1}\right)^{0 \nu} \left<\Delta h({\theta}_{\rm Tr})\Big|\frac{\partial h}{\partial{\lambda}^{\nu}}\right> \,,\label{eq:AlphaSys0}\\
    \sigma^2_\alpha &= \left(\Gamma^{-1}\right)^{00} =  \left[ \Gamma_{00} - \Gamma_{0i}\left(\gamma^{-1}\right)^{ij}\Gamma_{j0} \right]^{-1} \,. \label{eq:DeltaAlpha0}
\end{align}
where $\gamma_{ij}=\Gamma_{ij}$ is the lower-right block of the Fisher matrix, and in the final equality we have used the block-matrix inversion formula.

The optimal SNR, defined as $\rho(\lambda)=|h(\lambda)|$, is a convenient measure of the strength of the signal.
In order to investigate the scaling with the SNR, $\rho$, it will be convenient to separate it from the other parameters by defining the normalised model $\hat{h}=h/\rho$, with $|\hat{h}|=1$.
We also define the normalised Fisher matrix $\hat{\Gamma}_{\mu\nu}=\Gamma_{\mu\nu}/\rho^{2}$ and the normalised model error $\Delta \hat{h} = \Delta h/\rho$.

It will also be convenient to rescale the deviation parameter (i.e.\ redefine $\alpha\rightarrow \kappa \alpha$ where $\kappa$ is a constant) such that $\hat{\Gamma}_{00}\equiv|\partial\hat{h}/\partial\alpha|^{2}=1$. 
We are always free to perform such a rescaling and this does not interfere with our earlier choice of placing flat priors on all parameters.

We also assume that $\hat{\Gamma}_{0i}\equiv\big<\partial\hat{h}/\partial\alpha|\partial\hat{h}/\partial{\theta^i}\big>=0$ where $i\neq 0$, 
i.e. the deviation parameter induces waveform changes which are orthogonal to those arising from changes in all the GR parameters.
Although this is probably rarely true in practice, it is a conservative assumption in the sense that it makes the problem of waveform systematics as severe as possible by minimising the estimate for $\sigma_\alpha$ [see Eq.~(\ref{eq:DeltaAlpha0})], while keeping $\alpha_{\rm sys}$ fixed, thereby maximizing the chances that the model errors lead us to erroneously claim to have seen a deviation from GR.
It is this worst-case scenario which we choose to study here in order better understand when we need to worry about waveform systematics.

Under these simplifying assumptions and conventions, the statistical fluctuations in the deviation, given in Eq.~(\ref{eq:AlphaStat0}), are distributed as a Gaussian random variable,
$\alpha_{\rm stat}=z/\rho$ where $z\sim\mathcal{N}(0,1)$.
Furthermore, the expression for the standard deviation of the distribution, given in Eq.~(\ref{eq:DeltaAlpha0}), simplifies to $\sigma_\alpha=1/\rho$.
This just leaves the systematic offset in the deviation parameter in Eq.~(\ref{eq:AlphaSys0}), which can be written as
\begin{align}\label{eq:XYZ}
    \alpha_{\rm sys} &= \left(\hat{\Gamma}^{-1}\right)^{00} \left<\Delta \hat{h}(\theta_{\rm Tr})\bigg|\frac{\partial \hat{h}}{\partial\alpha}\right> =  |\Delta\hat{h}(\theta_{\rm Tr})|\cos\iota  \,, \end{align}
where the first equality follows from Eq.~(\ref{eq:AlphaSys0}) and our assumption that $\hat{\Gamma}_{0i}=0$, the second equality follows from our renormalization of $\alpha$ such that $|\partial\hat{h}/\partial\alpha|=1$, and $\iota$ is defined as the  angle between the signals $\Delta\hat{h}(\theta_{\rm Tr})$ and $\partial\hat{h}/\partial\alpha$.
The quantity $\iota$ has the interpretation of an angle if the signals (which are discretely sampled time series) are thought of as being very high-dimensional vectors in some signal space, $\mathcal{S}\backsimeq\mathbb{R}^{\rm high \,dim}$.
The angle $\iota$ encodes information on how the model error couples with the deviation parameter.
The worst-case scenario is when $|\cos\iota|$ is maximal and occurs when $\iota=0$ or $\pi$; therefore, we set $\cos\iota=\pm 1$ in the following. 
The norm of the model error, $|\Delta\hat{h}|$, is related to the mismatch which is commonly defined in GW applications as  (e.g.~\citealt{2008PhRvD..78l4020L})
\begin{align} \label{eq:MM_angle_phi}
    \mathcal{M}&=1-\frac{\big<\hat{h}+\Delta\hat{h} |\hat{h} \big>}{|\hat{h}||\hat{h}+\Delta\hat{h}|}
    = 1-\cos\phi \approx \frac{\phi^2}{2}\,, 
\end{align}
where $\phi$ is the generalised angle between the signals $\hat{h}$ and $\hat{h}+\Delta\hat{h}$.
Provided the $\Delta\hat{h}$ is small, the angle $\phi$ will also be small and is bounded above by $\phi < |\Delta\hat{h}|$.
The exact value of $\phi$ will depend on the details of the model error and can be considered to be quasi-random.
If the signal space dimensionality is large, and if $\Delta \hat{h}$ is a random vector, then the distribution of $\phi$-values will be peaked near the maximum value.
Therefore, we set $\phi = |\Delta\hat{h}|$ and, using the small angle approximation in Eq.~(\ref{eq:MM_angle_phi}), obtain
\begin{align} \label{eq:XYZ2}
    \mathcal{M}\approx\frac{ |\Delta\hat{h} |^{2}}{2}\,.
\end{align}
Finally, using Eq.~(\ref{eq:XYZ2}) to eliminate $\Delta\hat{h}$ from Eq.~(\ref{eq:XYZ}), the systematic error is $\alpha_{\rm sys} = \sqrt{2\mathcal{M}}\cos\iota$.
In summary, the 1D marginalised posterior on the GR deviation parameter $\alpha$ is given by Eq.~(\ref{eq:1Ddist}), with
\begin{align}
    \alpha_{\rm stat} &= \frac{z}{\rho} \,, \label{eq:AlphaStat}\\
    \alpha_{\rm sys} &= \sqrt{2\mathcal{M}} \cos\iota\,, \label{eq:AlphaSys}\\
    \sigma_\alpha &= \frac{1}{\rho} \,, \label{eq:DeltaAlpha}
\end{align}
where $z\sim\mathcal{N}(0,1)$ is a random number associated with the noise realisation and $\cos\iota = \pm 1$ is a random choice of sign associated with the model error, $\Delta \hat{h}$.

Note that the systematic offset does not scale with SNR. 
Therefore, there always exists a critical SNR above which we are in danger of erroneously claiming a deviation from GR.
When analyzing a single GW event for a deviation from GR, we are safe from the effects of model errors if $\alpha_{\rm sys} \ll \sigma_\alpha$.
From Eqs.~(\ref{eq:AlphaStat}-\ref{eq:DeltaAlpha}), we see that the average size statistical error equals the systematic error when $\rho= 1/\sqrt{2\mathcal{M}}$; therefore, we are safe from the effects of model errors if $\rho \ll 1/\sqrt{\mathcal{M}}$. 

Because the posterior in Eq.~(\ref{eq:like}) is Gaussian, it is possible to evaluate the Bayesian evidence integral analytically.
Doing so, and letting $k=\mathrm{dim}({\theta})$ [hence $\mathrm{dim}({\lambda})=k+1$)], gives
\begin{align}
    Z_{\rm nonGR} \equiv \int\mathrm{d}{\lambda}\;\mathcal{L}({\lambda}) = e^{c'}\sqrt{\frac{(2\pi)^{k+1}}{\mathrm{det}\,\Gamma_{\mu\nu}}} \,.
\end{align}

Because our waveform model $h(\alpha,\theta)$ is an extension of GR, it includes GR as a sub-model.
The GR submodel is the hypersurface $\alpha=0$ of the full model.
Using the same assumptions described above for the full model, the GR likelihood (and hence the posterior) on this hypersurface can be found from Eq.~(\ref{eq:like}) and is given by
\begin{align}
    \log\mathcal{L}_{\rm GR}({\theta}) =  c' &- \frac{1}{2}\Gamma_{00}\alpha_{\rm ML}^{2}  -\frac{1}{2}\left({\theta}-{\theta}_{\rm ML}\right)^i  \Gamma_{ij} \left({\theta}- {\theta}_{\rm ML}\right)^j\,, 
\end{align}
where $i,j\in\{1,2\ldots,k\}$ label the components of $\theta$.
The evidence for the GR sub-model can also be evaluated analytically and reads
\begin{align}
    Z_{\rm GR} &\equiv \int\mathrm{d}{\theta}\;\mathcal{L}_{\rm GR}({\theta})  =e^{c'-\Gamma_{\alpha\alpha}\alpha_{\rm ML}^{2}/2}\sqrt{\frac{(2\pi)^{k}}{\mathrm{det}\,\Gamma_{ij}}} \,. 
\end{align}
The odds ratio (or Bayes' factor) in favour of a deviation from GR is defined as
\begin{align}
\mathcal{B} \equiv \frac{\Pi}{A} \frac{Z_{\rm nonGR} }{ Z_{\rm GR} } \,,  
\end{align}
where $\Pi$ is the prior odds ratio in favour of a deviation from GR and $A=\alpha_{\rm max}-\alpha_{\rm min}$ is the prior range on $\alpha$ which must be included to account for the differing prior volumes between the models. 
Computing the Bayes' factor in this manner between nested models is known as the Savage-Dickey density ratio~\citep{dickey1971weighted}.
Under our conservative assumption that $\hat{\Gamma}_{0i}=0$, the Fisher matrix has a block diagonal structure and $\mathrm{det}\,\Gamma_{\mu\nu}=\Gamma_{00} \mathrm{det}\,\Gamma_{ij}$.
The Bayes' factor simplifies to 
\begin{align}
    \mathcal{B} &= \frac{\Pi}{A}\sqrt{\frac{2\pi}{\Gamma_{00}}}\exp\left(\frac{1}{2}\Gamma_{00}\alpha_{\rm ML}^{2}\right)\,.
\end{align}
Recalling that $\alpha_{\rm ML}=\alpha_{\rm stat}+\alpha_{\rm sys}$ and $\Gamma_{00}=\sigma_\alpha^{-2}$ and using the results in Eqs.~(\ref{eq:AlphaStat}-\ref{eq:DeltaAlpha}) gives 
\begin{align}\label{eq:log_odds}
    \log\mathcal{B} &= \log\left(\frac{\Pi}{A}\frac{\sqrt{2\pi}}{\rho}\right) + \frac{(z+\rho\sqrt{2\mathcal{M}}\cos\iota)^{2}}{2}\,.
\end{align}

Note that the first term in Eq.~(\ref{eq:log_odds}), known as the \emph{Occam penalty}, decays slowly with increasing SNR. 
However, the second term in Eq.~(\ref{eq:log_odds}) grows rapidly.
This reveals again, in another guise, the existence of a critical SNR above which we are in danger of erroneously claiming a deviation from GR due to model errors.
From the final term in Eq.~(\ref{eq:log_odds}), we again conclude that when analyzing individual GW events for deviations from GR we are safe from the effects of waveform systematics if $\rho\ll 1/\sqrt{\mathcal{M}}$.

%%%
\section{Linear signal analysis: event catalogs} \label{sec:catalog}

The previous section considered tests of GR with a single GW event and concluded that we expect our analysis to be robust against the effects of waveform systematic errors provided $\rho\ll1/\sqrt{\mathcal{M}}$. 
When we are in the situation that no single event in the catalog shows a clear deviation from GR, it is desirable to combine all the observed events to ``dig deeper'' and perform more stringent tests of GR.
This section extends the linearised analysis of the previous section and investigates the impact of waveform systematics on such catalog tests. 

A GW catalog contains $N$ events, indexed by $m\in\{1,2,\ldots,N\}$.
As described in the previous section, each event provides us with an independent measurement of the deviation parameter and
the likelihoods for these measurements, $P_{m}(\alpha)$, are all Gaussian of the form in Eq.~(\ref{eq:1Ddist}) with parameters given by Eqs.~(\ref{eq:AlphaStat0}-\ref{eq:DeltaAlpha0}), or Eqs.~(\ref{eq:AlphaStat}-\ref{eq:DeltaAlpha}). 
We replace $z\rightarrow z_{m}$, $\rho\rightarrow\rho_{m}$, $\cos\iota\rightarrow\cos\iota_m$ and $\mathcal{M}\rightarrow\mathcal{M}_m$ to distinguish different events.

There are different ways of combining the information from multiple events. 
Two particularly simple ways are (i) multiplying the 1D marginalised posteriors on $\alpha$ and (ii) multiplying the odds ratios~\citep{2019PhRvD..99l4044Z}. These approaches can been seen as two extrema of a more generic hierarchical-inference strategy  \citep{2019PhRvL.123l1101I}. In particular, the  former assumes that the deviation parameter takes the same value for each event in the catalog while latter assumes that the parameter takes an independent value for each event.
We consider each approach in turn.

\subsection{Multiplying likelihoods}

Under the assumption that the deviation takes the \emph{same} value in each event of the catalog, the combined posterior on the deviation parameter is given by the product of the independent likelihoods in each of the $N$ events;
\begin{align} \label{eq:combined_posterior_alpha}
    P_{\rm same}(\alpha) &= \prod_{m=1}^{N}P_{m}(\alpha)= \frac{1}{\sqrt{2\pi}}\,{\exp\left[-\frac{(\alpha-\alpha^{\rm same}_{\rm ML})^2}{2(\sigma_\alpha^{\rm same})^2}\right]}\,. 
\end{align}
The product of several Gaussian distributions with known means and variances is another Gaussian, the mean and variance of which are given by
\begin{align} \label{eq:norm_combine_means}
    \alpha_{\rm ML}^{\rm same}&= \big(\sigma_\alpha^{\rm same}\big)^{2} \sum_{m=1}^{N}\frac{\alpha_{\mathrm{ML},m}}{\sigma_{\alpha,m}^{2}} \,,\\
    \sigma_{\alpha}^{\rm same} &= \left(\sum_{m=1}^{N}\sigma_{\alpha,m}^{-2}\right)^{-1/2} \,. \label{eq:norm_combine_sigmas}
\end{align}

Using the combined catalog posterior on the deviation parameter in Eq.~(\ref{eq:combined_posterior_alpha}), we can now compute combined Bayes' factor in favour of a deviation from GR. 
This is computed using the Savage-Dickey density ratio and reads (see e.g.~\citealt{sivia2006data})
\begin{align} \label{eq:logB_same_result}
    \mathcal{B}_{\rm same} = \frac{\Pi}{A} \sqrt{2\pi}\sigma_\alpha^{\rm same}\exp\left[
    \frac{1}{2}\left(\frac{\alpha^{\rm same}_{\rm ML}}{\sigma_\alpha^{\rm same}}\right)^{2}
    \right] \,.
\end{align}

\subsection{Multiplying Bayes' factors} \label{subsec:MB}

Each catalog event provides some evidence for or against a deviation from GR which is quantified by the Bayes' factor $\mathcal{B}_{i}$ in Eq.~(\ref{eq:log_odds}).
Under the assumption that the deviation takes independent values in each event, the combined Bayes' factor in favour of a deviation from GR is given by the product 
\begin{align} \label{eq:logB_diff_result}
    \mathcal{B}_{\rm diff}=\prod_{m=1}^{N}\mathcal{B}_{m} \,.
\end{align}

\section{Simple event catalogs}\label{sec:toy}

In this section we perform Monte-Carlo simulations of highly simplified, mock GW catalogs. It is assumed throughout that GR is the correct description of nature, but that our GR waveforms contain modelling errors. 
The purpose of these simulations is to understand under what situations model errors might lead us to mistakenly think that we have observed a deviation from GR.

For simplicity, in this section it is assumed that all events in the catalog have the same SNR, $\rho$.
It is also assumed that all events have the same amount of modelling error and that this leads to a mismatch value of $\mathcal{M}=10^{-3}$ in each event.
Finally, in this section the effects of instrumental noise are also neglected; i.e. it is assumed that the specific noise realisation in each observed GW event is $n=0$, which corresponds to setting $z=0$ for each event.
All of these assumptions are relaxed in Sec.~\ref{sec:real} where more realistic catalogs are considered.

We simulate a mock catalog by first choosing the number of events to be considered, $N$.
We then choose the value of the SNR, $\rho$. 
All that remains is to choose the value of $\cos\iota = \pm 1$ for every event; this is done in two ways described below.

We can then compute the evidence in favour of a deviation from GR either under the assumption that the deviation parameter takes the same value for each event [$\mathcal{B}_{\rm same}$, see Eq.~(\ref{eq:logB_same_result}); i.e. multiplying likelihoods] or else under the assumption that the deviation takes independent values in each event [$\mathcal{B}_{\rm diff}$, see Eq.~(\ref{eq:logB_diff_result}); i.e. multiplying Bayes' factors].
We consider these two cases in turn.

\subsection{Multiplying likelihoods}

Each individual event, labelled by $m\in\{1,2,\ldots,N\}$, gives a measurement of the deviation parameter. 
Under the assumptions described in Sec.~\ref{sec:linear_sig_analysis}, and neglecting the statistical fluctuations due to the noise, the likelihood on $\alpha$ from this measurement is a 1D Gaussian with a mean $\alpha_{\mathrm{ML},m} = \alpha_{\mathrm{sys},m}$  given by Eq.~(\ref{eq:AlphaSys}) and a standard deviation $\sigma_{\alpha,m}$ given by Eq.~(\ref{eq:DeltaAlpha}).

Multiplying these likelihood functions together gives a single, combined catalog measurement of the deviation parameter.
The likelihood from this combined measurement is also a 1D Gaussian with a mean and standard deviation given by Eqs.~(\ref{eq:norm_combine_means}) and (\ref{eq:norm_combine_sigmas}) respectively.
These expressions simplify further to give
\begin{align}
    \alpha^{\rm same}_{\rm ML} &= \frac{\sqrt{2 \mathcal{M}}}{N} \left(\sum_{m=1}^{N}\cos\iota_m\right) \,, \\
    \sigma_\alpha^{\rm same} &= \frac{1}{\sqrt{N}\rho}\,.
\end{align}

The Bayes' factor in favour of a deviation from GR that comes from this combined catalog measurement of $\alpha$ was derived in Eq.~(\ref{eq:logB_same_result}) and simplifies further here to give
\begin{align} \label{eq:new_deriv_B}
    \log\mathcal{B}_{\rm same} &= \log\left(\frac{\Pi}{A}\frac{\sqrt{2\pi}}{\sqrt{N}\rho}\right) + \frac{\mathcal{M}\rho^{2}}{N} \left(\sum_{m=1}^{N}\cos\iota_m\right)^2 \,.
\end{align}

\begin{figure*}[t]
\centering
  \includegraphics[width=\textwidth]{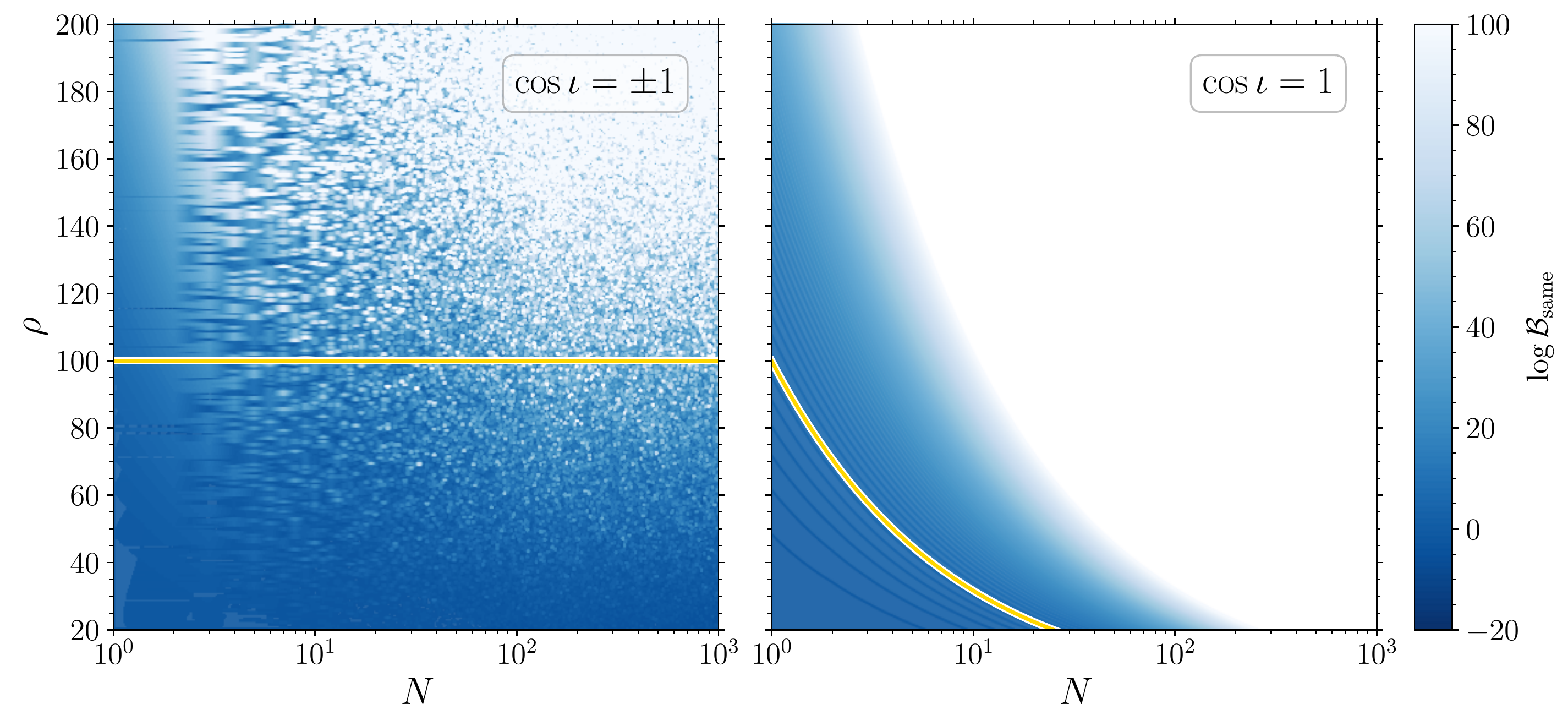}
\caption{ \label{fig:linearized_results}
    The log Bayes' factor, $\log\mathcal{B}_{\rm same}$, in favour of a deviation from GR under the assumptions that the deviation parameter takes the same value in all events. 
    These results were obtained for the highly simplified mock GW event catalogs described in Sec.~\ref{sec:toy}; in particular, this assumes that each of the $N$ events has the same SNR, $\rho$, and the same model mismatch error, $\mathcal{M}=10^{-3}$, and we neglect noise fluctuations by setting $z=0$.
    Left panel: the model errors are equally likely to favour positive or negative $\alpha$ ($\cos\iota=\pm 1$ randomly in all events); in this case the model errors do not accumulate strongly when combining the catalog events.
    The speckled pattern comes from the random, Monte-Carlo choices for $\cos\iota$ in each event and would tend to average out if we simulated multiple catalog realisations.
    Right panel: the model errors always favour positive $\alpha$ ($\cos\iota=1$ in all events); in this case the model errors accumulate rapidly as the number of events increases and $\mathcal{B}_{\rm same}$ increases with $N$. 
    The yellow lines shows the analytic prediction of the threshold $\log\mathcal{B}_{\rm same}\approx 10$ above which model errors might cause us to erroneously claim to have detected a deviation from GR;
    the horizontal line in the left hand comes from Eq.~(\ref{eq:new_deriv_55}), while the line in the right hand figure comes from Eq.~(\ref{eq:new_deriv_2}).
    In both panels the analytic predictions for the threshold follow the contours of the heat map.}
\end{figure*}

The heat maps in Fig.~\ref{fig:linearized_results} show the numerical, Monte-Carlo results for the Bayes' factor $\mathcal B_{\rm same}$ under two possible scenarios. The left panel of Fig.~\ref{fig:linearized_results} illustrates a case where the model errors differ among events such that they are equally likely to favour positive and negative value for $\alpha$.
We mimic this scenario by randomly selecting either $\cos\iota=1$ or $\cos\iota=-1$ for each event.
The right panel of Fig.~\ref{fig:linearized_results} illustrates the case where the model errors are such that they always tend to favour a deviation of $\alpha$ with the same sign.
We mimic this scenario by always choosing $\cos\iota=+1$ in every event.
In reality, the situation is likely to be somewhere in between these two extreme possibilities.
The real distribution of $\cos\iota$ will depend on the astrophysical population of sources and any detection biases.
Unless we are very unlucky, the modelling error is unlikely to always resemble exactly the same type of deviation from GR. 
However, because we analyze all GW events using the same waveform model the modelling errors are also not independent between events.
For simplicity, the results in Fig.~\ref{fig:linearized_results} are scaled to $\Pi=A=1$.

It is possible to understand analytically the distinctly different scaling of $\log\mathcal{B}_{\rm same}$ observed in the two panels of Fig.~\ref{fig:linearized_results}.
Firstly, we consider case where the model errors are such that they always tend to favour a deviation of $\alpha$ with the same sign (right panel).
In this case $\cos\iota_m=+1$ for every event and we simply have that $\sum_{m}\cos\iota = N$. 
For large catalogs, the expression for the Bayes' factor in Eq.~(\ref{eq:new_deriv_B}) now becomes 
\begin{align} \label{eq:new_deriv_2}
    \log\mathcal{B}_{\rm same} &\approx \mathcal{M}N\rho^{2} \,.
\end{align}
The logarithm term is neglected as we are mainly interested in the limiting behavior for large $N$ and $\rho$.

\begin{figure*}[t]
\centering
  \includegraphics[width=0.58\linewidth]{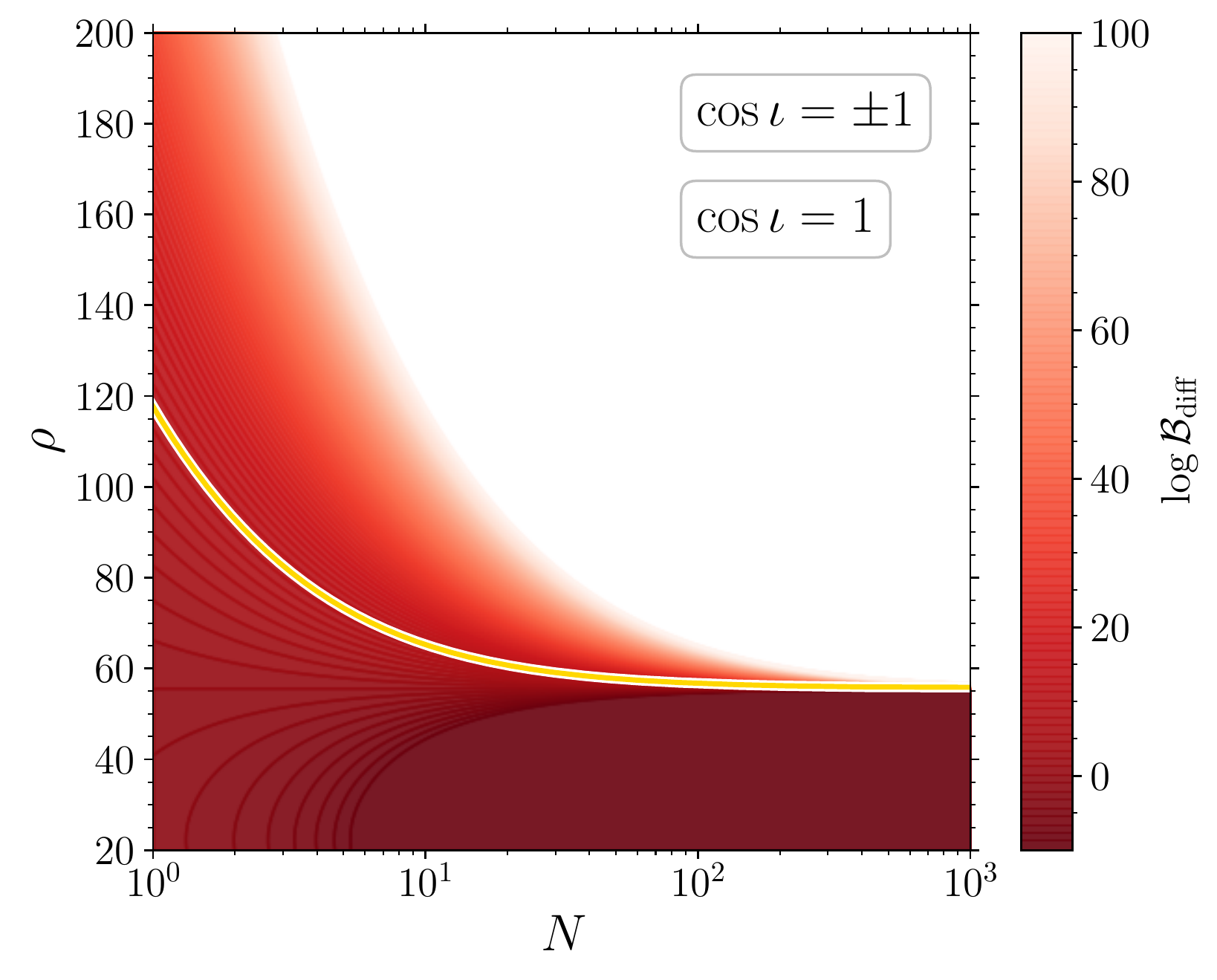}
\caption{ \label{fig:linearized_results_two}
    The log Bayes' factor, $\log\mathcal{B}_{\rm diff}$, in favour of a deviation from GR under the assumption that the deviation parameter takes independent values in each event. 
    These results were obtained for the highly simplified mock GW event catalogs described in Sec.~\ref{sec:toy}.
    In this situation, the two cases where the model errors equally favour positive and negative $\alpha$ ($\cos\iota=\pm 1$) and where they always favour positive $\alpha$ ($\cos\iota=1$) give identical results. 
    The yellow lines shows the analytic prediction in Eq.~(\ref{eq:31}) of the threshold $\log\mathcal{B}_{\rm diff}=10$ above which model errors might cause us to erroneously claim to have detected a deviation from GR;
    this prediction closely follows the contours of the heat map.
    Here, the model errors accumulate only if the SNR in each individual event is above a critical value, $\rho_{i}\gtrsim \rho_{*} = 55.68$ [see Eq.~(\ref{eq:snr_c})].
    If the individual event SNRs are below this critical value then the Bayes' factor actually decreases with increasing catalog size leading us to (correctly) favour the GR hypothesis.
    }
\end{figure*}

If we choose an arbitrary threshold Bayes' factor (say, $\mathcal{B}_{\rm threshold}=e^{10}$) above which we will claim to have seen evidence for a deviation from GR, then rearranging Eq.~(\ref{eq:new_deriv_2}) gives an expression for the threshold SNR as a function of catalog size.
This is plotted as the yellow curve in the right panel of Fig.~\ref{fig:linearized_results} where it can be seen to follow the contours of the heat map.
We see that even if $\rho\ll 1/\sqrt{\mathcal{M}}$, and our waveform model is comfortably good enough to analyse each event individually, there always exists a critical catalog size about which the Bayes' factor in favour of a deviation from GR exceeds any threshold.
In this case, as the catalog size increases there is a growing danger of erroneously claiming to detect a deviation from GR due to the model error.

Secondly, we consider the case where the model errors differ among events such that they are equally likely to favour positive and negative value for $\alpha$.
This scenario was mimicked in our toy model by choosing $\cos\iota_m=\pm1$ randomly.
Therefore, the term $\sum_{m}\cos\iota$ is a new random variable, and in the limit of large catalog size (i.e. as $N\rightarrow\infty$) the central limit theorem implies that this will be normally distributed as $\sum_{m}\cos\iota\sim\mathcal{N}(0,\sqrt{N})$.
It follows that the combination $(\sum_{m}\cos\iota)^2/N$ appearing in Eq.~(\ref{eq:new_deriv_B}) is now distributed as a $\chi^2$ random variable with 1 degree of freedom and has an expectation value of 1. 
Therefore, the expectation value for the Bayes' factor in Eq.~(\ref{eq:new_deriv_B}) becomes
\begin{align} \label{eq:new_deriv_55}
    \log\mathcal{B}_{\rm same} &\approx \mathcal{M}\rho^{2} \,.
\end{align}
Again, we neglect the logarithm term as it is unimportant in the limit of large $\rho$.
This expression can be rearranged to find the threshold SNR above which $\mathcal{B}_{\rm same}$ exceeds the threshold; this is plotted as the horizontal yellow line in the left panel of Fig.~\ref{fig:linearized_results}.
Note the very different scaling from that in Eq.~(\ref{eq:new_deriv_2}); in this case, the model errors do not accumulate as the catalog size increases and the danger of erroneously claiming a deviation from GR does not increase with $N$.

\subsection{Multiplying Bayes' factors}

Figure~\ref{fig:linearized_results_two} shows the results of another Monte-Carlo analysis, this time combining the catalog events under the assumption that the GR deviation parameter takes independent values in each event.
As discussed in Sec.~\ref{subsec:MB}, this corresponds to multiplying together the Bayes' factors for each individual catalog event in order to obtain the combined $\mathcal{B}_{\rm diff}$ catalog Bayes' factor in favour of a GR deviation.

Again, we consider a simplified GW catalog containing $N$ events each at the same SNR, $\rho$.
As before, we further assume that each event has the same mismatch, $\mathcal{M}=10^{-3}$, due to modelling errors and we neglect the statistical fluctuations due to noise by setting $z=0$ for each event.
Again, we could consider both a scenario where the model errors equally favour positive and negative $\alpha$ (i.e. randomly selecting $
\cos\iota_m\pm1$) and a scenario where the model errors always favour positive $\alpha$ (i.e. always setting $\cos\iota_m=1$). However, in this worstcase scenario, these two possibilities give identical results (inspecting Eq.~(\ref{eq:log_odds}) we see that, when setting $z=0$, the individual event Bayes' factor $\mathcal{B}_{m}$ depends only on $\cos\iota_m^2$).

The heat map in Fig.~\ref{fig:linearized_results_two} shows the numerical, Monte-Carlo results for the Bayes' factor $\mathcal{B}_{\rm diff}$.
As before, it is possible to understand analytically the observed scaling of $\log\mathcal{B}_{\rm diff}$.
The Bayes' factor for each individual event is given by Eq.~(\ref{eq:log_odds}) (with $z=0$, as we are neglecting statistical noise fluctuations in this section).
The combined log Bayes' factor $\mathcal{B}_{\rm diff}$ is simply the sum of the individual log Bayes' factors and is given by
\begin{align} \label{eq:31}
    \log\mathcal{B}_{\rm diff} &\approx N\left[\log\left(\frac{\Pi}{A}\frac{\sqrt{2\pi}}{\rho}\right) + \mathcal{M} \rho^{2}\right]  \,.
\end{align}

This expression can be rearranged to find the SNR at which the Bayes' factor exceeds the threshold for claiming  evidence for a deviation from GR.
This predicted threshold SNR is plotted as a yellow line in Fig.~\ref{fig:linearized_results_two} for the choice $\mathcal{B}_{\rm threshold} =e^{10}$.

In this case we see a qualitatively new behavior as the catalog size, $N$, increases.
Whenever a new event is added to the catalog, there is a competition between the model error [second term in Eq.~(\ref{eq:31})] which tends to increase the Bayes' factor in favour of a deviation from GR and the Occam penalty [first term in Eq.~(\ref{eq:31})] which tends to do the opposite. Which effect ends up winning depends on the SNR. There exists a critical SNR, $\rho_{*}$, above which the Bayes' factor increases with $N$ and this is given by the solution to 
\begin{align} \label{eq:snr_c}
    \log\left(\frac{A\rho_{*}}{\sqrt{2\pi }\Pi}\right) = \mathcal{M}\rho_{*}^{2} \,,
\end{align}
which in our example where $\Pi=A=1$ and $\mathcal{M}=10^{-3}$ is $\rho_{*} = 55.68$.
Below this critical SNR we are safe from model systematics and the Bayes' factor in favour of a deviation from GR actually decreases as the catalog grows. In the large-$N$ limit and within the assumption of this model, this implies that evidence against GR grows (is suppressed) in  catalogs made of events with SNR $\rho>\rho_{*}$ ($\rho<\rho_{*}$).

\section{More realistic event catalogs}\label{sec:real}

The GW catalogs considered in the previous section were rather unrealistic. 
The SNR of each event was the same, the mismatch was the same for every waveform, and the statistical fluctuations due to individual noise realisations was ignored. 
In this section we relax these assumptions and perform Monte-Carlo simulations of more realistic catalogs.

\begin{figure*}[t]
\centering
  \includegraphics[width=\textwidth]{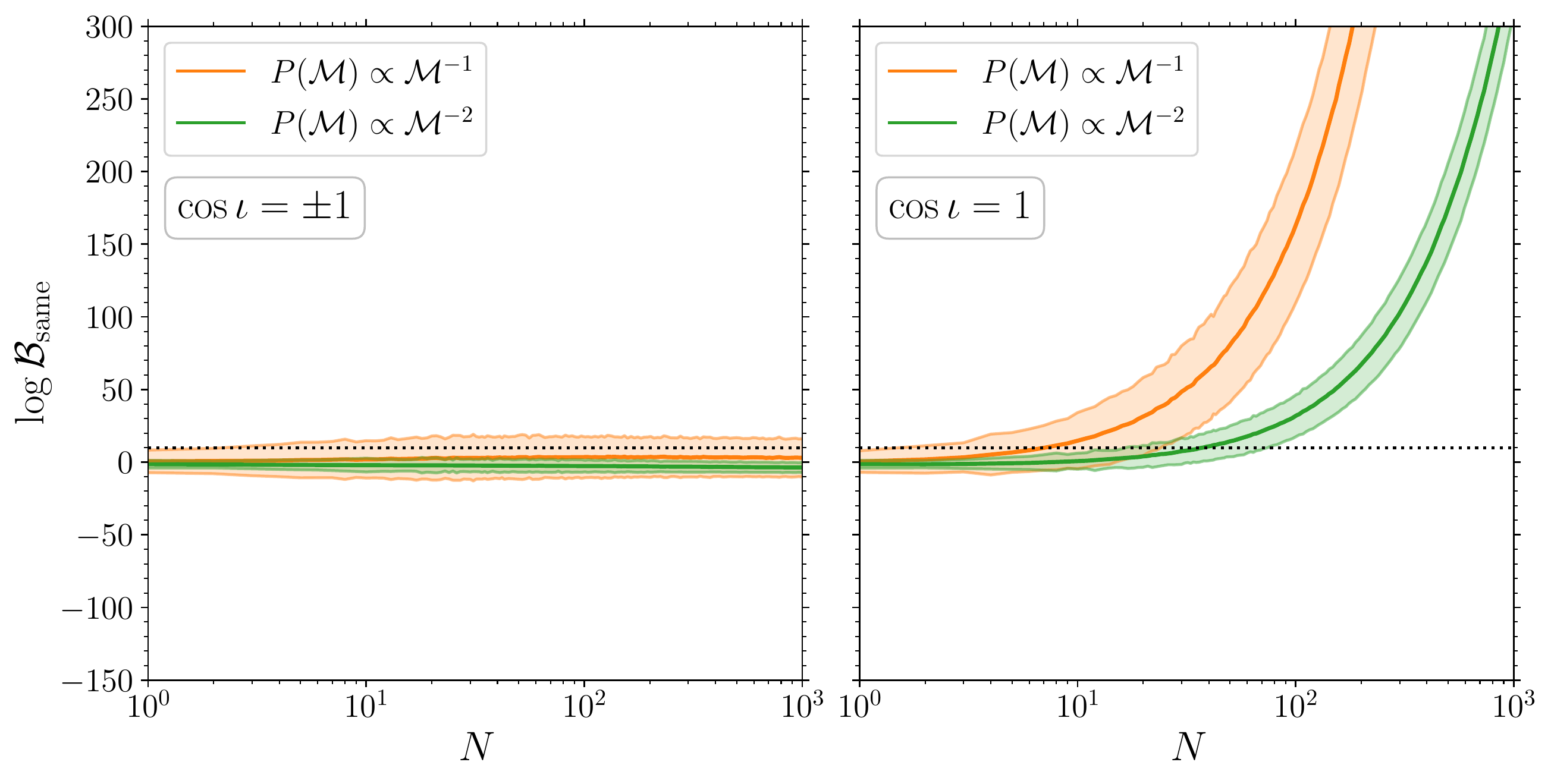}
\caption{ \label{fig:linearized_results_real}
    The log Bayes' factor, $\log\mathcal{B}_{\rm same}$, in favour of a deviation from GR under the assumption that the deviation parameter takes the same value in all events. 
    The solid lines indicates the mean value of $\log\mathcal{B}_{\rm diff}$ obtained from $10^4$ realisations of the more realistic simulated catalogs described in Sec.~\ref{sec:real} while the shaded region between the two paler lines indicates the $\pm 1\sigma$ spread in this set of simulated catalogs.
    The dashed horizontal line denotes the threshold $\log\mathcal{B}_{\rm same}=10$: above this line there is a risk that model errors cause us to incorrectly claim a deviation from GR, while below this line we correctly conclude that GR is favoured.
    Left panel: the model errors are equally likely to favour positive or negative $\alpha$ ($\cos\iota=\pm 1$ randomly in all events); in this case the model errors do not accumulate strongly when combining the catalog events. 
    Right panel: the model errors always favour positive $\alpha$ ($\cos\iota=1$ in all events); in this case the model errors accumulate rapidly as the number of events increases and the evidence for a deviation from GR grows with the size of the catalog.
    Depending on the distribution of the model errors, misleading evidence for a deviation from GR can appear with catalogs with as few as $\approx 10$ events above the minimum SNR of $\rho>20$.}
\end{figure*}

We simulate catalogs of $N$ events where the SNR of individual events are drawn from a $P(\rho)\propto\rho^{-4}$ distribution, which is the expected distribution for a population of sources in a Euclidean universe with no cosmological evolution in the merger rate \citep{2011CQGra..28l5023S,2014arXiv1409.0522C}.
The lower (upper) cutoffs in the SNR distribution where chosen to be $\rho_{\rm low}=20$ ($\rho_{\rm high}=200$).
Our results are somewhat sensitive to the lower cutoff of the SNR distribution; the value of 20 used here is larger than the usual LIGO/Virgo detection threshold $\rho\to 8$ because: (i) we do not want to invalidate the assumptions behind the linearised analysis which are only expected to hold for large SNR, and (ii) it is reasonable to expect that delicate analyzes such as tests of GR will only be performed on a subset of loud events. This was done, for example, in the recent analysis by \cite{2020arXiv201014529T}
where none of the marginal triggers with false alarm rate $>10^{-3}\,\mathrm{yr}^{-1}$ were investigated. 

\begin{figure*}[t]
\centering
  \includegraphics[width=\textwidth]{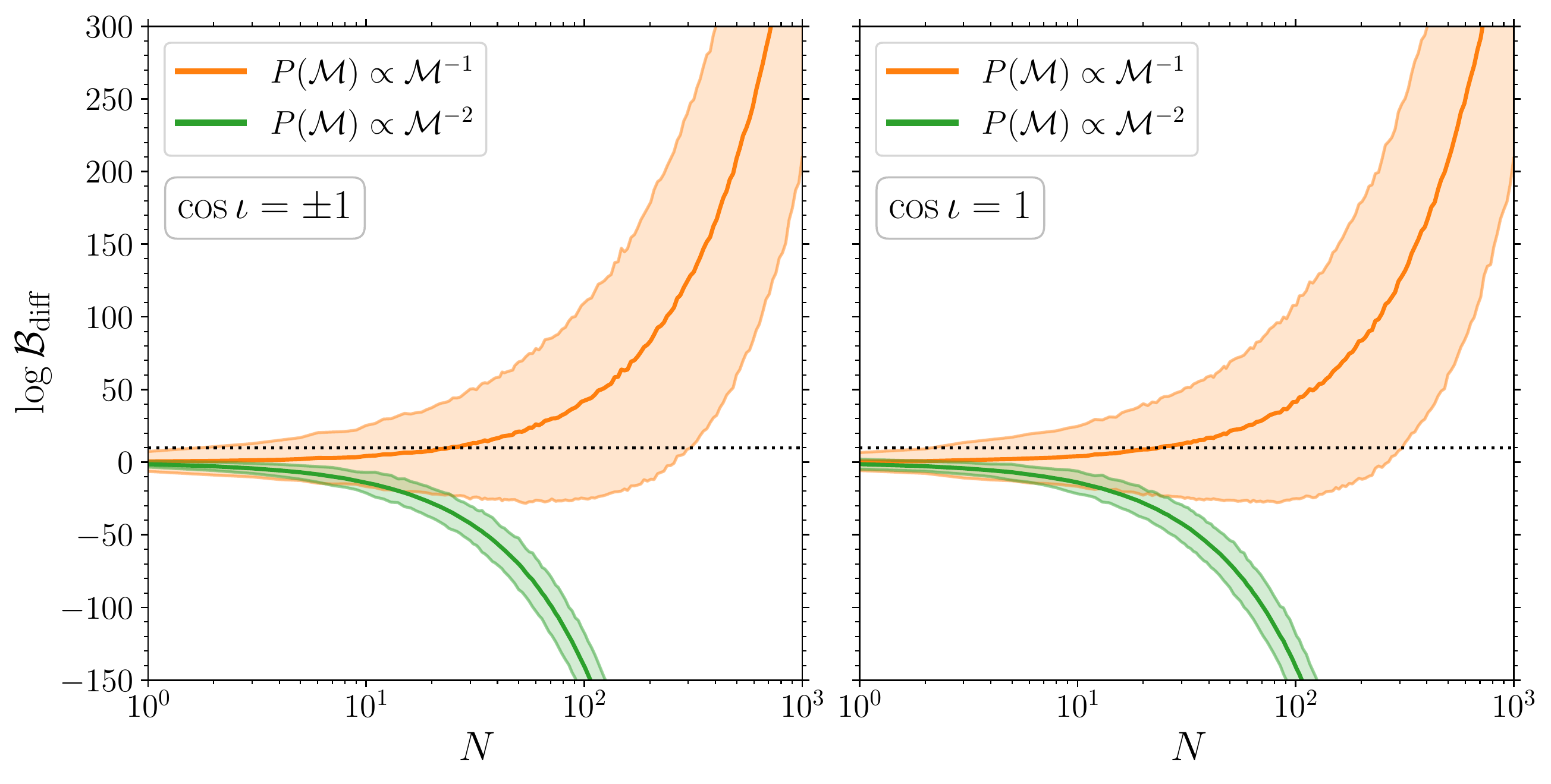}
\caption{ \label{fig:linearized_results_real_two}
    The log Bayes' factor, $\log\mathcal{B}_{\rm diff}$, in favour of a deviation from GR under the assumption that the deviation parameter takes independent values in all events.
    The solid lines indicate the mean value of $\log\mathcal{B}_{\rm diff}$ obtained from $10^4$ realisations of the simulated catalog while the shaded region between the two paler lines indicates the $\pm1\sigma$ spread in this set of simulated catalogs.
    The dashed horizontal line denotes the threshold $\log\mathcal{B}_{\rm same}=10$.
    Left panel: the model errors are equally likely to favour positive or negative $\alpha$ ($\cos\iota=\pm 1$ randomly in all events).
    Right panel: the model errors always favour positive $\alpha$ ($\cos\iota=1$ in all events).
    In both cases we see the evidence for a deviation from GR grows rapidly with the size of the catalog if our waveform models are bad (i.e. $P(\mathcal{M})\propto \mathcal{M}^{-1}$) but decreases rapidly if our models are good (i.e. $P(\mathcal{M})\propto \mathcal{M}^{-2}$).
    In the worst case, misleading evidence for a deviation from GR can appear with catalogs containing as few as $\approx 30$ events above the minimum SNR of $\rho>20$.
    }
\end{figure*}

Instead of fixing the mismatch at a single value $\mathcal{M}=10^{-3}$ for all events, we now allow the mismatch to differ between events by drawing this from a distribution with lower (upper) cutoffs of $\mathcal{M}_{\rm low}=10^{-4}$ ($\mathcal{M}_{\rm high}=10^{-2}$).
The choice of these cutoffs is roughly motivated by the accuracy of existing models and the results from Fig.~13 of \cite{2017PhRvD..95j4023B}.
The shape distribution of $\mathcal{M}$ between these limits is difficult to predict as it will depend on the waveform models used, on where in parameter space this model perform best/worst, and on the distribution of the event properties such as mass ratio and spins presented to us nature in the catalog.
All of these are difficult to predict.
However, given existing observations, we expect most events will be nearly equal mass and with low spins \citep{2020arXiv201014533T} where our waveform models perform relatively well (although a small number of more exotic events should be expected).
Therefore, the distribution of $\mathcal{M}$ will be skewed towards low values.
Here, we consider two possibilities: a bad case $P(\mathcal{M})\propto\mathcal{M}^{-1}$ and a good case $P(\mathcal{M})\propto\mathcal{M}^{-2}$.

The log Bayes' factors obtained from the catalogs assuming the deviation takes the same value in every event (i.e. $\mathcal{B}_{\rm same}$; multiplying likelihoods) are shown in Fig.~\ref{fig:linearized_results_real}.
We consider the same two cases for $\cos\iota$ as in Sec.~\ref{sec:toy}. 
In the left hand panel we see that the Bayes' factor does not scale strongly with the size of the catalog; this agrees with the results in the left panel of Fig.~\ref{fig:linearized_results} obtained using the simpler catalogs.
In the right hand panel, we see that the Bayes' factor in favour of a deviation from GR increases rapidly with the size of the catalog. This is also in agreement with the results in the right panel of Fig.~\ref{fig:linearized_results} obtained using the simpler catalogs.

The results for the log Bayes' factors obtained assuming that the deviation parameter takes independent values in each event  (i.e. $\mathcal{B}_{\rm diff}$; multiplying the individual Bayes' factors) are shown in Fig.~\ref{fig:linearized_results_real_two}.
Again, we consider both $\cos\iota=\pm 1$ and $\cos\iota=1$.
We see very similar behaviour in both cases (consistent with the identical results found in Sec.~\ref{sec:toy}).
In both panels we see that the Bayes' factor scales strongly with the size of the catalog but that it can either increase or decrease depending on the distribution of the mismatches. This behavior can be understood from the results in Fig.~\ref{fig:linearized_results_two} obtained using the simpler catalogs.
If $P(\mathcal{M})\propto\mathcal{M}^{-2}$, most events have very small mismatches and therefore have $\rho<\rho_{*}$ (i.e. below the yellow line in Fig.~\ref{fig:linearized_results_two}) and, as the catalog size increases, the increasing Occam penalty dominates over the effect of the model error and GR is favoured.
On the other hand, if $P(\mathcal{M})\propto\mathcal{M}^{-1}$, more events have larger mismatches and $\rho>\rho_{*}$ (i.e. above the yellow line in Fig.~\ref{fig:linearized_results_two}) and, as the catalog size increases, the accumulating model errors overcome the Occam penalty and a deviation from GR is favoured.

From the results in Figs.~\ref{fig:linearized_results_real} and \ref{fig:linearized_results_real_two}, in four of the eight ``realistic'' scenarios considered here the misleading evidence in favour of a deviation from GR due to the modelling errors accumulates rapidly with increasing catalog size. 
This occurs even if the waveform model is good enough to safely analyse each event in the catalog individually.
These results highlight the potentially insidious effects of waveform systematics when performing testing of GR with catalogs of GW events.

%%%
\section{Discussion} \label{sec:discussion}

Developing waveform models is a challenging task that inevitably involves some approximations, simplifications and modelling errors. 
These include truncating post-Newtonian series at some high order, neglecting certain physical effects (e.g. tidal terms, subdominant spin effects and orbital eccentricity) and the finite accuracy in numerical-relativity simulations. 
If the resulting models are interpreted at face value, these systematic offsets can mimic the effect of new physics beyond GR.  

This is a rather generic effect that has long been known about at the level of individual events.
In this paper, we show how this extends to the case when a catalog of events is analysed for signs of a deviation from GR.
Using a simple, linearised analysis we have studied whether and how fast the modelling errors accumulate and have shown that it depends on:
\begin{enumerate}
\item the alignment of the model errors with the particular deviation from GR under consideration (i.e.\ does the modelling error always tend push $\alpha$ in one direction, or does it vary across parameter space and tend to average out across many different events); 
\item how the catalog events are combined to give a test of GR (i.e. whether the deviation is assumed to take the same value in each event [multiplying likelihoods], independent values [multiplying Bayes' factors], or some intermediate case);
\item the distribution of waveform modelling errors (i.e. mismatches $\mathcal{M}$) across catalog events, which in turn depends on the waveform models used and the location of new events in parameter space.
\end{enumerate} 

Furthermore, our idealised calculation shows that this is a rather urgent problem.
Erroneous evidence for new physics from waveform systematics \emph{might} occur with as few as $10-30$ events at SNR $\gtrsim 20$. 
Although this is a conservative estimate and reflects the worst-case scenario (cf. Sec.~\ref{sec:linear_sig_analysis}), it is dangerously close to the size of current catalogs.

Going forward, our Fisher-like analysis needs to be backed up by injection and recovery campaigns. This will address more realistically the details of how current waveform models perform when used for a selection of parameterised tests of GR on catalogs of various sizes.

\section*{Acknowledgements}
%{\small
We thank
Antoine Klein,
Geraint Pratten,
Elinore Roebber,
Patricia Schmidt,
Lucy Thomas,
and
Alberto Vecchio 
for discussions.
D.G. is supported by European Union's H2020 ERC Starting Grant No. 945155--GWmining, Leverhulme Trust Grant No. RPG-2019-350, and Royal Society Grant No. RGS-R2-202004. 
Computational work was performed on the University of Birmingham BlueBEAR cluster.
%}

\bibliographystyle{elsarticle-harv}
{\small
\setlength{\bibsep}{2pt plus 0.3ex}
\bibliography{refs}
}

\end{document}